\documentclass[
aps,
reprint,
amsmath,amssymb,prx
]{revtex4-2}

\usepackage{blindtext}
\usepackage{graphicx}% Include figure files
\graphicspath{{./Figures}}
\usepackage{dcolumn} % Align table columns on decimal point
\usepackage{bm} % bold math
\usepackage{braket}
\usepackage{xcolor}
\usepackage{dsfont}
\usepackage{comment}
\usepackage{tikz}
\usepackage{zx-calculus}
\usepackage{booktabs}
\usepackage{amsmath}
\usepackage{subcaption}
\usepackage{wrapfig}

\usepackage{xparse}

\NewDocumentCommand{\tens}{e{_^}}{%
  \mathbin{\mathop{\otimes}\displaylimits
    \IfValueT{#1}{_{#1}}
    \IfValueT{#2}{^{#2}}
  }%
}

\def \beq {\begin{eqnarray}}
\def \eeq {\end{eqnarray}}

\begin{document}

\title{Comment on arXiv:2501.17230 and 2502.00103\\ ``Phonon-mediated electron attraction in SrTiO3 via the generalized Fröhlich and deformation potential mechanisms'' \\
and\\
``Theory of ab initio downfolding with arbitrary range electron-phonon
coupling''}
\date{February 2024}

\author{Jonathan Ruhman}
\affiliation{Department of Physics, Bar-Ilan University, 52900, Ramat Gan, Israel}

\begin{abstract}
This comment critically examines the claims made in arXiv papers [2501.17230] and [2502.00103], which  argue that a multiplicity of  polar optical phonons can generate a long-range attractive interaction via a ``generalized Fröhlich coupling.'' I identify a fundamental flaw in their derivation, showing that their result relies on an unphysical assumption—specifically, neglecting the intermode Coulomb interactions between different polar optical phonons. By restoring these missing interactions I show the screened Coulomb interaction is always repulsive in the static limit. 
\end{abstract}

\maketitle
\section{Introduction}
One of the reasons superconductivity is a remarkable phenomenon regards the fact that it involves  electronic pairing, which goes against any classical logic or intuition. For example, if we treat the crystal as a classical dielectric medium, it will not ``overscreen'' the interaction, resulting in a long-range Coulomb attraction. That would imply a  negative dielectric constant and an unbounded negative electrostatic energy 
\[ U ={1\over 2} \int d^3 r\, \varepsilon_0 |\boldsymbol E|^2\,,\]
which would imply the crystal is unstable.

All known pairing mechanisms include some quantum ingredient, i.e. a process that has no classical counterpart.
One example is the  Anderson-Morel mechanism~\cite{PhysRev.125.1263}, which shows that dressing the pairing vertex with virtual excitations to high energy (the Gor'kov ladder),  affects only the instantaneous part of the interaction. Thus, in an interaction with  instantaneous repulsion and  retarded attraction, they screen only the former. Another example, is the Kohn-Luttinger model~\cite{PhysRevLett.15.524}, in which a $2k_F$ non-analyticity in the renormalized interaction renders it attractive in high angular momentum channels. The latter is also inherently quantum because it requires a Fermi surface, which is a purely quantum entity.  Such mechanisms do not have  classical intuitive  pictures describing how repulsion is overcome in superconductors.

Recently, the authors of Refs.~\cite{tubman2025phononmediatedelectronattractionsrtio3,tubman2025theoryabinitiodownfolding} have claimed to discover a new pairing mechanism, which defies this rule.  Namely, they claim that a multiplicity of  polar phonons (e.g. in lightly doped SrTiO$_3$) is sufficient to generate a long-ranged attractive interaction in a transparent crystal. Using what they call a ``generalized Fro\"hlich coupling''  they derive the static limit of the screened Coulomb repulsion   
\begin{equation}\label{eq:wrong} 
V_{C}(\boldsymbol r - \boldsymbol r') = \left(1 -\sum_\nu {\omega_{LO,\nu}^2-\omega_{TO,\nu}^2\over \omega_{LO,\nu}^2}\right){e^2\over \varepsilon{_\infty} |\boldsymbol r - \boldsymbol r'|}\,, 
\end{equation}
where $\omega_{TO,\nu}$ and $\omega_{LO,\nu}$ are the transverse and longitudinal optical frequencies of the polar optic mode $\nu$, which  they derive from DFT. Clearly, this relation implies a long-ranged attractive interaction  if $\sum_\nu {\omega_{LO,\nu}^2-\omega_{TO,\nu}^2\over \omega_{LO,\nu}^2}>1$, which also implies a negative dielectric constant~\footnote{It should be noted that free charge carriers were not considered in this calculation, and if correct, it holds for pristine SrTiO$_3$. 
It should also be noted that the screening by polar optical modes is fully captured within  RPA and classical screening theory. Therefore, it would be a real paradigm shift if such a classical mechanism would suffice for an attractive interaction. }.

Notwithstanding, in this comment I show that Eq.~\eqref{eq:wrong} is the result of an unphysical assumption. In particular, it can be reproduced by assuming each one of the polar phonons interacts via long-ranged Coulomb interaction with itself but not with the other modes. When this fallacy is corrected,  the standard Lyddane–Sachs–Teller (LST) relation for the static Coulomb repulsion in a crystal with multiple polar optic phonons~\cite{PhysRevB.22.5501} is restored
\begin{equation}\label{eq:right} V_{C}(\boldsymbol r - \boldsymbol r') = \left(\prod_\nu {\omega_{TO,\nu}^2 \over \tilde\omega_{LO,\nu}^2}\right){e^2\over \varepsilon{_\infty} |\boldsymbol r - \boldsymbol r'|}\,, \end{equation}
where I added a tilde to $\tilde \omega_{LO,\nu}$ to denote the renormalization in the longitudinal frequencies due to the  inter-mode interaction. 

It should be noted that the idea that a multiplicity of polar optical phonons can lead to a negative dielectric constant was proposed in Ref.~\cite{gor2016phonon} and then withdrawn in Ref.~\cite{gor2017back} (also see Ref.~\cite{ruhman2016superconductivity}).

\section{summary of results}
To derive the dielectric constant I assume a model with multiple polar optical phonon modes denoted by $\boldsymbol u_\nu$. The propagators for the longitudinal  modes is expanded close to the zone center  
\begin{align}  D_0^{-1}(i\omega,\boldsymbol q) = {\rho_\nu } \left[ \omega^2+c_L^2 q^2 +\omega_{TO,\nu}^2 \right]\,.\nonumber\end{align}
where $\rho_\nu$ is the mass density of mode $\nu$ and $\omega_{TO,\nu}$ is the  optical phonon frequency prior to the LO-TO splitting.
Because the optical phonons are polar their divergence induces a \emph{bound charge}
\begin{equation}
\rho_{\textrm{bound}}^\nu = Q_\nu \nabla\cdot \boldsymbol u_\nu\,.
\end{equation}
This bound charge interacts with other polar modes and electron charge.  
In appendix \ref{app:dielectric}  I derive an expression for the dielectric constant resulting from this interaction
\begin{equation}\label{eq:eps_main}
\varepsilon(i\omega,\boldsymbol q) = \varepsilon_\infty \left(  1-{4\pi  \over \varepsilon_\infty} \boldsymbol Q^T \tilde D(i\omega,\boldsymbol q)\boldsymbol Q \right)^{-1}\,,
\end{equation}
where 
\begin{align}\label{eq:D_main} \tilde D_{\nu\nu'}^{-1}&(i\omega,\boldsymbol q) =\\& {\rho_\nu } \left[ \omega^2 + c_{L,\nu}^2 q^2+\omega_{LO,\nu}^2 \right]+{4\pi Q_{\nu} Q_{\nu'}\over \varepsilon_{\infty} } (1-\delta_{\nu \nu'})\,.\nonumber\end{align}
is the renormalized phonon propagator in mode space, $\boldsymbol Q^T = (Q_1,Q_2,\ldots)$ is a vector of the parameters $Q_\nu$ and we define the frequencies
\begin{equation}\label{eq:wLO}\omega_{LO,\nu}^2 \equiv \omega_{TO,\nu}^2 + {4\pi Q_\nu^2 \over \varepsilon_\infty \rho_\nu}\,.\end{equation}
It is important to note that these frequencies \textbf{are not the longitudinal optical frequencies}, as they will be renormalized by the off-diagonal term in Eq.~\eqref{eq:D_main}, which stems from the inter mode Coulomb repulsion.

In appendix \ref{app:wrong} I derive equation \eqref{eq:wrong} by simply omitting the off-diagonal terms in Eq.~\eqref{eq:D_main}. The physical implication of such an omission is that the bound charges of different polar modes do not interact with one another, which is obviously unphsyical. 
The fact that I reproduce the exact same equation as in Refs.~\cite{tubman2025phononmediatedelectronattractionsrtio3,tubman2025theoryabinitiodownfolding} implies they make the same unphysical assumption. 
On the other hand, when I properly take these off-diagonal terms into account the dielectric constant assumes the form Eq.~\eqref{eq:right}  and the universe breathes a sigh of relief. Notably, the dielectric constant remains positive in the static limit and on the entire imaginary axis (see appendix \ref{app:2modecase} and \ref{app:multimode}).

\begin{figure}[hbt!]
    \centering
    \includegraphics[width=0.85\linewidth]{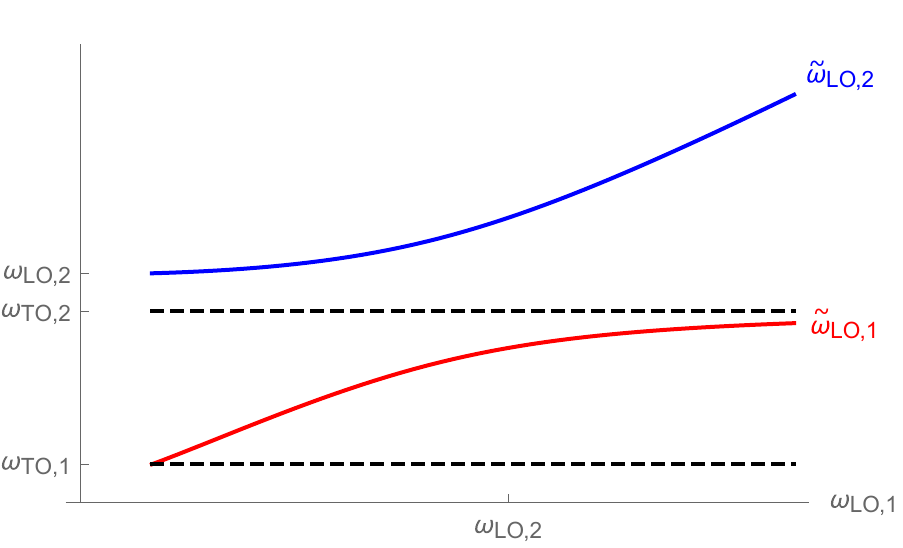}
    \caption{ The longitudinal optical frequencies,  $\tilde \omega_{LO,\nu}$, in the case a two-mode model plotted as a function of $\omega_{LO,1}$, which is a function of model parameter $Q_1$  Eq.~\eqref{eq:wLO}, which quantifies the polarity of mode $\nu =1$. }
    \label{fig:slc}
\end{figure}

The LO mode frequencies are  the zeros of the dielectric constant Eq.~\eqref{eq:eps_main}. I  denote them with a tilde $\tilde \omega_{LO,\nu}$  to distinguish them from the frequencies \eqref{eq:wLO}. They  are plotted in Fig.~\ref{fig:slc} as a function of the bare parameter $\omega_{LO,1}$ for fixed $\omega_{LO,2}$ and $\omega_{TO,1}<\omega_{TO,2}$ (see  appendix \ref{app:2modecase} for more details ). Tuning $\omega_{LO,1}$ is equivalent to tuning $Q_1$, through Eq.~\eqref{eq:wLO}. 
For $Q_1 = 0$, $\tilde \omega_{LO,1} = \omega_{TO,1}$ and $\tilde \omega_{LO,2} = \omega_{LO,2}$. Then $\tilde \omega_{LO,1}$ grows as we increase the polarity of the first mode $Q_1$. However, due to the off-diagonal terms in Eq.~\ref{eq:D_main} there is an avoided crossing with $\tilde \omega_{LO,2}$ and $\tilde \omega_{LO,1}$  is bounded from above by $\omega_{TO,2}$.

\section{single mode limit}
A good way to test the expressions Eq.~\eqref{eq:wrong} and \eqref{eq:right} is to take them to the extreme limit  $Q_1\to\infty$ (i.e. the polarity of one of the modes goes to infinity). In that case, we have a huge LO frequency, which screens all other modes and thus, the single mode case is restored. 
To that end, let me consider the two-mode case plotted in Fig.~\ref{fig:slc}. 
I take $Q_1\to \infty$, which implies $\omega_{LO,1}\to\infty$ [see Eq.~\eqref{eq:wLO}], I get $\tilde\omega_{LO,2} = \omega_{LO,1}$ and $\tilde\omega_{LO,1} = \omega_{TO,2}$ [see Eq.~\eqref{eq:tildewLO}]. Plugging this in Eq.~\eqref{eq:right} gives 
\[V_C(\boldsymbol r - \boldsymbol r') ={\omega_{TO,1}^2 \over \omega_{LO,1}^2}{e^2\over \varepsilon{_\infty} |\boldsymbol r - \boldsymbol r'|} \]
which is nothing but the single mode LST relation.  
On the other hand, if I plug this limit in Eq.~\eqref{eq:wrong} I get 
\[V_C(\boldsymbol r - \boldsymbol r') =-\left(3-{\omega_{TO,1}^2 \over \omega_{LO,1}^2}-{\omega_{TO,2}^2 \over \omega_{LO,2}^2}\right){e^2\over \varepsilon{_\infty} |\boldsymbol r - \boldsymbol r'|} \,.\]
which is attractive and stronger than the bare Coulomb repulsion in magnitude! Note that even if I plug the correct (renormalized) frequencies $\tilde \omega_{LO,\nu}$ into Eq.~\eqref{eq:wrong} (in the limit of $Q_1\to \infty$) I  get an attractive interaction
\[V_C(\boldsymbol r - \boldsymbol r') =-\left(3-{\omega_{TO,1}^2 \over \omega_{TO,2}^2}-{\omega_{TO,2}^2 \over \omega_{LO,1}^2}\right){e^2\over \varepsilon{_\infty} |\boldsymbol r - \boldsymbol r'|} \,.\]
Both expressions above do not respect the LST relation.  This shows the interaction in Eq.~\eqref{eq:wrong} is inconsistent with standard lore and lacks the important physics of inter-mode screening, even if the correct LO and TO frequencies are used. 
 
% \section{Correspondence with the authors} Full disclosure, I had a long email exchnage with the authors of Refs.~\cite{tubman2025phononmediatedelectronattractionsrtio3,tubman2025theoryabinitiodownfolding}. During this exchange they made the claim that the dielectric constant corresponding to their interaction is consistent with the LST relation. As far as I could understand the claim was that their expression Eq.~\eqref{eq:wrong} is only relevant at small but finite $q$ and that exactly at $q=0$ the dielectric function jumps from a negative value to the LST relation. I do not understand this claim, but i strongly appose it: The prefactor in front of the long-ranged electronic interaction $e^2/|\boldsymbol r - \boldsymbol r'|$ is the static dielectric constant at $q\to 0$ by definition. 

% They also sent me a bunch of references discussing the possibility of the dielectric constant becoming negative~\cite{RevModPhys.53.81}. However, all concrete  examples are either at finite frequency or include some form of free charge carriers that can stabilize the system. Moreover, they completely disregard the fact that I reproduce their result with basically the same starting point and an unphsyical assumption. 

\section{Conclusions}
I conclude that  Eq.~\eqref{eq:wrong} results from an unphysical assumption; that the bound charge generated by different polar phonon modes do not interact with each other. Once these modes are allowed to interact through a long-ranged Coulomb interaction the dielectric constant takes its standard LST form. 
Thus, I conclude that the Coulomb attraction derived in Refs.~\cite{tubman2025phononmediatedelectronattractionsrtio3,tubman2025theoryabinitiodownfolding} is based on a fallacy.  

\emph{Acknowledgments.-} JR was funded by the Simons foundation and the ISF under grant No. 915/24.

\bibliography{bibi}

\appendix
\section{Model}\label{app:model}
Following the methods in Ref.~\cite{kozii2019superconductivity} I consider a field-theoretic description of the crystal close to zone center in imaginary time (Matsubara frequency), such that the partition function is given by 
\[ \mathcal Z = \int D[\phi,\boldsymbol u_\nu,\bar \psi, \psi ]e^{-\sum_{\omega,\boldsymbol q}\mathcal L[\phi,\boldsymbol u_\nu,\bar \psi, \psi]} \,,\]
where $\phi$ is the Coulomb field, $\boldsymbol u_{\nu}$ are the polar optical modes and $\psi$ is the electronic field. The Lagrangian is given by 
\begin{widetext}
\begin{align}\label{eq:action}
\mathcal L[\phi,\boldsymbol u_\nu,\bar \psi, \psi] = {\epsilon_\infty q^2 \over 8\pi} |\phi_q|^2 + {1\over 2}\sum_\nu (u_\nu ^i)^* [D_{\nu}^{-1}(i\omega,\boldsymbol q)]_{ij}u_\nu^j +\bar\psi G^{-1}\psi +ie n_{q}\phi_{-q}+ \sum_{\nu}Q_\nu [\boldsymbol q\cdot \boldsymbol u_\nu( q)]\phi_{- q} \,,
\end{align}
where the first term is the electrostatic energy in the absence of the phonons and the second and third terms are the free Lagrangian's of the phonons and electrons, respectively (i.e. $D$ and $G$ are the respective Green's functions). The last two term are the coupling between of the electrons and phonons to  the Coulomb field, respectively.  Namely, $n_q = {1\over V} \sum_k \bar\psi_{k+q}\psi_k$ is the electronic density at wave-vector $q$ and $Q_{\nu}$ is the proportionality factor between the divergence of the optical phonon, $\boldsymbol u_\nu$, and the \emph{bound charge density}.
Thus, $Q_{\nu}$ has units of charge$\times$ionic density.  

Close to the zone center the Green's function of the optical modes assumes the form 
\[[D_{\nu}^{-1}(i\omega,\boldsymbol q)]_{ij} = {\rho_\nu } \left[ \left(\omega^2 +\omega_{TO,\nu}^2\right)\delta_{ij} + c_{L,\nu}^2q_i q_j +c_{T,\nu}^2 (q^2 \delta_{ij}-q_i q_j) \right] \,,\]
where $\rho_\nu$ is the ionic mass density of mode $\nu$, $\omega_{TO,\nu}$ is the transverse optical frequency, which is currently also the longitudinal frequency because I have not taken into account the LO-TO splitting yet. $c_{L,\nu}$ and $c_{T,\nu}$ are the longitudinal and transverse velocities. 

For convenience I  denote the longitudinal modes by
$$\varphi_\nu(\boldsymbol q) \equiv \hat {\boldsymbol q} \cdot \boldsymbol u_\nu(\boldsymbol q)\quad \to \quad \varphi_\nu(-\boldsymbol q) = - \varphi_\nu^*(\boldsymbol q)$$
and omit the transverse fields hereafter since they are decoupled from the Coulomb field.
Furthermore, the free electron action takes no role in the calculation I am about to perform, so I can omit it as well. 
Overall, the Lagrangian then simplifies to 
\begin{align}\label{eq:action2}
\mathcal L = {\epsilon_\infty q^2 \over 8\pi} |\phi_q|^2 + {1\over 2}\sum_\nu \varphi_\nu^*  D_{\nu}^{-1}(i\omega,\boldsymbol q) \varphi_\nu +i\,e\, n_{q}\phi_{-q}+ q\, \boldsymbol Q^T  \cdot \boldsymbol \varphi_{q}\,\phi_{- q}\,,
\end{align}
where 
\[D_{\nu}^{-1}(i\omega,\boldsymbol q)= {\rho_\nu } \left[ \omega^2 +\omega_{TO,\nu}^2 + c_{L,\nu}^2 q^2\right] \]
and $\boldsymbol Q^T = (Q_1,Q_2,\ldots)$ and $\boldsymbol \varphi^T = (\varphi_1,\varphi_2,\ldots)$ are vectors in the space of modes $\nu$. 

\section{Derivation of the dielectric constant}\label{app:dielectric}
The derivation of the dielectric constant follows through two steps. In the first step I integrate out the Coulomb field $\phi$. then in the next step I integrate out the longitudinal phonon. I am left with a term proportional to $|n_q|^2$ which is nothing but the screened Coulomb repulsion from which the dielectric constant is readily obtained. 

It is worth noting that because all couplings are linear and the bosonic action is Gaussian, each integration out of a mode amounts to completing a square, nothing more. It is also worth noting that this implies the screening I am taking into account is completely classical and within the RPA.  

Thus, I start with the first step and integrate out $\phi$. I get
\begin{align}\label{eq:action3}
\mathcal L =  {1\over 2}\sum_\nu \varphi_\nu^*   D_{\nu}^{-1}(i\omega,\boldsymbol q) \varphi_\nu+ {2\pi  \over \varepsilon_\infty q^2}\left(e^2\, |n_{q}|^2- i2e q\, \boldsymbol Q^T  \cdot \boldsymbol \varphi_{q}n_{-q}+q^2\sum_{\nu \nu'}Q_\nu Q_{\nu'}\varphi_\nu^* \varphi_{\nu'}\right)\,,
\end{align}
The first term is just the bare Coulomb repulsion in the absence of the phonons. The second term is the Fr\"ohlich coupling of the polar phonons to the electrons. The last term  is quadratic in the fields $\phi_\nu$, therefore I can  define the Green's function as a matrix in mode space
\begin{equation}\label{eq:D} \tilde D_{\nu\nu'}^{-1}(i\omega,\boldsymbol q) = \left( D_{\nu}^{-1}(i\omega,\boldsymbol q) + {4\pi Q_\nu^2 \over \varepsilon_\infty} \right)\delta_{\nu\nu'} +{4\pi  \over \varepsilon_\infty}\left(Q_{\nu}Q_{\nu'}-Q_{\nu}^2 \delta_{\nu \nu'} \right) \,.\end{equation}
\end{widetext}
Here I divided the Green's function into diagonal and off-diagonal parts. The latter represents the Coulomb interaction between the bound charges associated with the polar modes. Below I show that by neglecting these off-diagonal elements the erroneous equation \eqref{eq:wrong} in Ref.~\cite{tubman2025phononmediatedelectronattractionsrtio3} is obtained. 

However, before I get to that I first perform the last step and integrate out the longitudinal phonons. This leads to action 
\begin{equation}
\mathcal L = {1\over 2}{4\pi e^2 \over \varepsilon_\infty q^2 }\left(  1-{4\pi  \over \varepsilon_\infty} \boldsymbol Q^T \tilde D(i\omega,\boldsymbol q)\boldsymbol Q \right)|n_q|^2
\end{equation}
which leads to the expression
\begin{equation}\label{eq:eps}
\varepsilon(i\omega,\boldsymbol q) = \varepsilon_\infty \left(  1-{4\pi  \over \varepsilon_\infty} \boldsymbol Q^T \tilde D(i\omega,\boldsymbol q)\boldsymbol Q \right)^{-1}\,.
\end{equation}

\section{The single mode case}
I can verify this result is physical by considering the limit of a single mode. This is done by nullifying all $Q_\nu$ except $Q_1$. Then I have 
\[ -{4\pi e^2 \over \varepsilon_\infty} \boldsymbol Q^T \tilde D(i\omega,\boldsymbol q)\boldsymbol Q =- {\omega_{LO,1}^2 - \omega_{TO,1}^2 \over \omega^2 +\omega_{LO,1}^2 +c_{L,1}^2 q^2 } \]
where I used the definition 
\begin{equation}\label{eq:bare_wLO}
    \omega_{LO,\nu}^2 \equiv \omega_{TO,\nu}^2 + {4\pi Q_\nu^2 / \rho_\nu \varepsilon_\infty }
\end{equation}
Then, I plug this in Eq.~\eqref{eq:eps} and obtain 
\[ \varepsilon(i\omega,\boldsymbol q) = \varepsilon_\infty {\omega^2 + c_{L,1}^2 q^2 + \omega_{LO,1}^2 \over \omega^2 + c_{L,1}^2 q^2 + \omega_{TO,1}^2}\,, \]
\\
which reproduces the LST relation in the static homogeneous limit and agrees with the results of Ref.~\cite{kozii2019superconductivity}. 

\[\]

\section{The two mode case}\label{app:2modecase}
In the case of two modes the Green's function \eqref{eq:D} assumes the form (for brevity I take $q=0$, although it is trivial to generalize to finite $q$)
\begin{widetext}
\[ D(i\omega,0) ={1\over \rho_1 \rho_2 \left( \omega^2+\omega_{LO,2}^2  \right) \left( \omega^2\omega_{LO,1}^2 + \right) - {16\pi^2 Q_1^2 Q_2^2 / \varepsilon_\infty^2}} \begin{pmatrix} \rho_2 \left( \omega^2+\omega_{LO,2}^2  \right) & {4\pi Q_1 Q_2}/\varepsilon_\infty \\
{4\pi Q_1 Q_2}/\varepsilon_\infty & \rho_1 \left( \omega^2+\omega_{LO,1}^2 \right)
\end{pmatrix}\]
Using the above matrix and the relation $Q_{\nu} = \sqrt{(\omega_{LO,\nu}^2-\omega_{TO,\nu}^2)\rho_\nu\varepsilon_\infty/4\pi}$ in Eq.~\eqref{eq:eps} I get
\begin{equation}
 \varepsilon(\omega,0) = \varepsilon_\infty  {(\omega^2 + \tilde\omega_{LO,1}^2)(\omega^2 + \tilde\omega_{LO,2}^2)\over (\omega^2 + \omega_{TO,1}^2)(\omega^2 + \omega_{TO,1}^2)}\,,
\end{equation}
where I added a tilde to the frequencies $\tilde\omega_{LO,\nu}$ to denote the renormalized longitudinal frequency 
\begin{align}\label{eq:tildewLO}
\tilde\omega_{LO,1}^2 = {1\over 2}\left( \omega_{LO,1}^2+\omega_{LO,1}^2-\sqrt{(\omega_{LO,1}^2+\omega_{LO,1}^2)^2+4\omega_{TO,1}^2 \omega_{TO,2}^2 -4 \omega_{TO,1}^2 \omega_{LO,2}^2 -4 \omega_{TO,2}^2 \omega_{LO,1}^2}\right) \\
\tilde\omega_{LO,2}^2 = {1\over 2}\left( \omega_{LO,1}^2+\omega_{LO,1}^2+\sqrt{(\omega_{LO,1}^2+\omega_{LO,1}^2)^2+4\omega_{TO,1}^2 \omega_{TO,2}^2 -4 \omega_{TO,1}^2 \omega_{LO,2}^2 -4 \omega_{TO,2}^2 \omega_{LO,1}^2}\right)\nonumber 
\end{align}
\end{widetext}
These two frequencies are plotted in Fig.~\ref{fig:slc} as a function of the parameter $\omega_{LO,1}$, which is defined through Eq.~\eqref{eq:bare_wLO}. 
Notably, the smaller one of the two frequencies, $\tilde \omega_{LO,1}$, is always bounded between the two transverse modes $\omega_{TO,1}$ and $\omega_{TO,2}$. This is a direct result of the screening.

\section{The multi mode case}\label{app:multimode}
The more modes I add to the problem the more cumbersome will the analytic expression become. However, it is straightforward to test numerically that the structure of the dielectric constant generalizes to the form of Eq.~\eqref{eq:right}. Thus, the homogeneous and static dielectric constant is always positive on the imaginary frequency axis, as it should be according to electrostatic theory. 

\section{Deriving the Erroneous Eq. (1) by Deliberately Neglecting off-Diagonal Terms }\label{app:wrong}
Finally, I turn to the error in Refs.~\cite{tubman2025phononmediatedelectronattractionsrtio3,tubman2025theoryabinitiodownfolding}. 
I use Eq.~\eqref{eq:eps} to show that Eq.~\eqref{eq:wrong} is obtained through an \textbf{unphysical assumption}. Namely, I neglect the off-diagonal terms in Eq.~\eqref{eq:D}, in which case it becomes diagonal and is readily inverted. This yields the following identity 
\[ -{4\pi e^2 \over \varepsilon_\infty} \boldsymbol Q^T \tilde D(i\omega,\boldsymbol q)\boldsymbol Q = -\sum_{\nu}{\omega_{LO,\nu}^2-\omega_{TO,\nu}^2 \over \omega^2 + c_{L,\nu}^2 q^2 +\omega_{LO,\nu}^2 } \]
The inverse dielectric constant is obtained by plugging this expression in Eq.~\eqref{eq:eps}.
In the limit of $q\to 0$ and $\omega \to 0$ this yields
\[ {1\over \varepsilon_0} = {1\over \varepsilon_\infty} \left(1 - \sum_{\nu}{\omega_{LO,\nu}^2 - \omega_{TO,\nu}^2 \over \omega_{LO,\nu}^2} \right)\,, \]
which is identical to Eq.~\eqref{eq:wrong}. 
Thus, I reproduce the result of Refs.~\cite{tubman2025phononmediatedelectronattractionsrtio3,tubman2025theoryabinitiodownfolding} by assuming that the bound charge of each mode interacts  with the electronic charge but not with the bound charge associated with other polar modes.

\end{document}